\documentclass[]{iopart}
\usepackage{color}
\usepackage{graphicx}
\usepackage[T1]{fontenc}
\usepackage{ae,aecompl}
\usepackage[english]{babel}

\newlength{\figa}
\begin{document}


\title{Configurational entropy of Wigner crystals}
\author{A Radzvilavi\v{c}ius and E Anisimovas}
\address{Department of Theoretical Physics, Vilnius University,
Saul\.{e}tekio al.~9, LT-10222 Vilnius, Lithuania}

\date{17 November 2010}

\begin{abstract}
We present a theoretical study of classical Wigner crystals in two- and 
three-dimensional isotropic parabolic traps aiming at understanding and 
quantifying the configurational uncertainty due to the presence of multiple 
stable configurations. Strongly interacting systems of classical charged 
particles confined in traps are known to form regular structures. The number 
of distinct arrangements grows very rapidly with the number of particles,
many of these arrangements have quite low occurrence probabilities and 
often the lowest-energy structure is not the most probable one. We perform 
numerical simulations on systems containing up to 100 particles interacting 
through Coulomb and Yukawa forces, and show that the total number of metastable 
configurations is not a well defined and representative quantity. Instead, 
we propose to rely on the configurational entropy as a robust and objective 
measure of uncertainty. The configurational entropy can be understood as the 
logarithm of the effective number of states; it is insensitive to the presence 
of overlooked low-probability states and can be reliably determined even 
within a limited time of a simulation or an experiment.
\end{abstract}


\submitto{\JPCM}

\maketitle


\section {Introduction}

Under the conditions of low disorder and temperature, a dilute electron gas
forms an ordered structure known as the Wigner crystal \cite{Wigner,Crandall}.
This term is also used to describe any other ordered system of strongly
interacting particles, usually confined by an external potential. Typical
experimental realizations of such structures are electrons on the surface of
liquid helium \cite{Crandall,Rousseau}, electrons in semiconductor quantum
dots \cite{Filinov}, cooled ions in various kinds of traps \cite{Dubin},
vortices in superfluids \cite{Saari} and micrometric charged 
spheres in low-temperature dusty plasmas \cite{Bonitz review}.

Ordered structures of two-dimensional (2D) and three-dimensional (3D) charged
particle systems confined by an isotropic parabolic potential have been
investigated in numerous recent studies. The configurations of stationary
states have been both observed experimentally
\cite{Saint Jean,Arp,Bonitz Struct} and calculated theoretically
\cite{Bedanov,Ying-Ju,Kong}. It turns out that in most cases besides the
unique ground-state configuration corresponding to the global energy minimum,
a number of \emph{metastable} configurations corresponding to various local
energy minima are present.

\begin{figure}[ht]
\centering
\includegraphics[width=\figa]{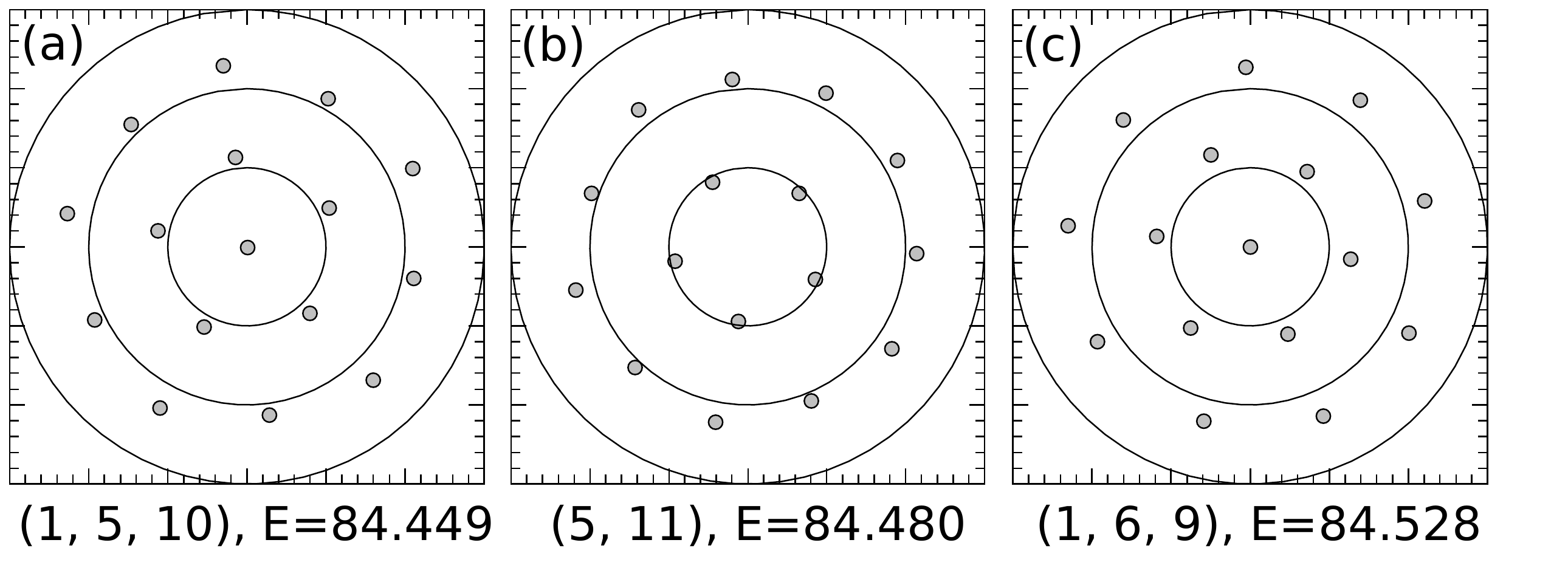}
\caption{Ground (a) and metastable (b, c) states of a 16-particle 2D Coulomb
cluster. The concentric circles mark equipotential lines of the parabolic trap.
Numbers in parentheses list the numbers of particles in each shell starting 
with the innermost one. Note that the energy differences are quite small.}
\label{2D-16}
\end{figure}

\begin{figure}[ht]
\centering
\includegraphics[width=\figa]{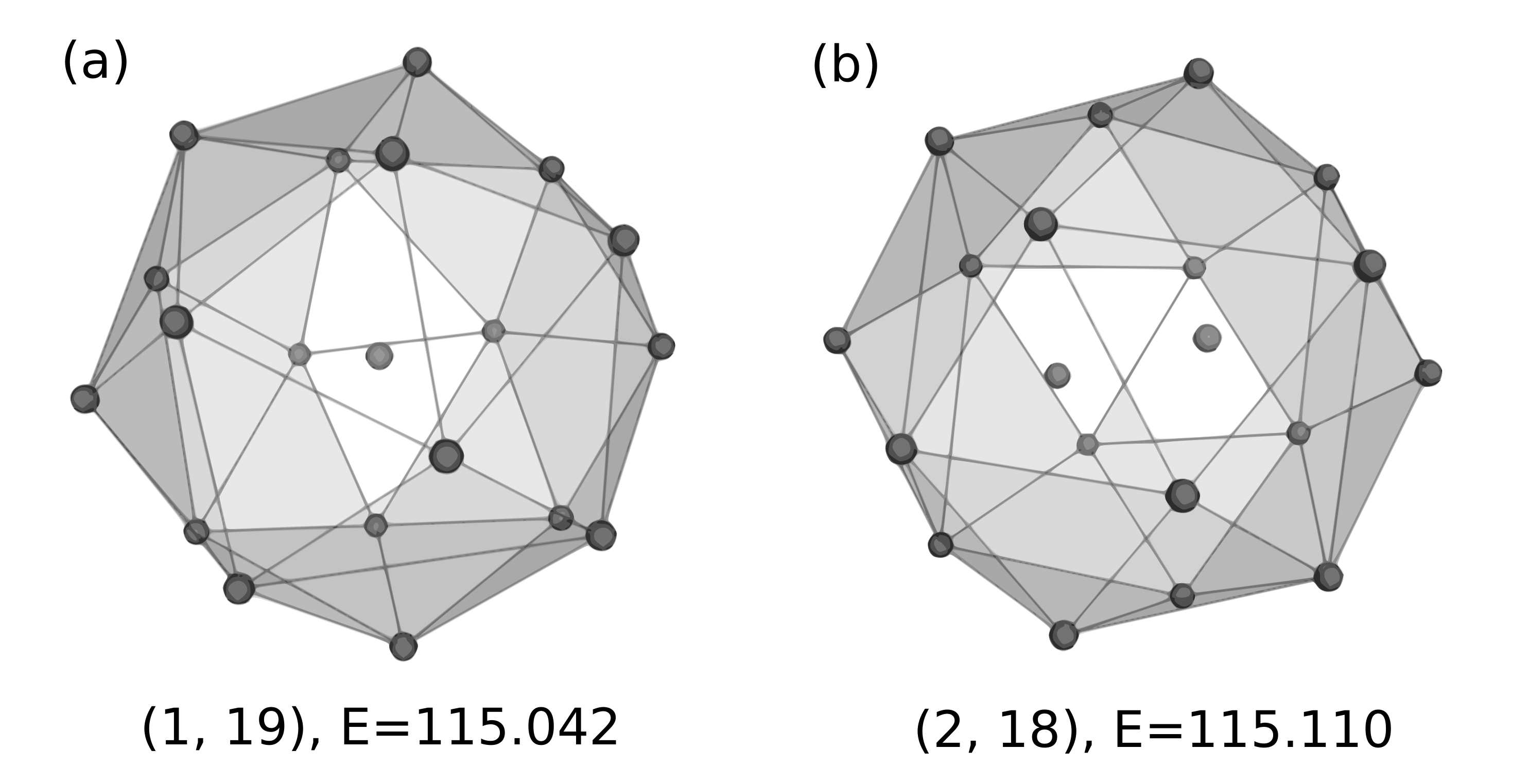}
\caption{Ground (a) and metastable (b) states of a 20-particle 3D Coulomb
cluster. The inner shell contains just one (a) or two (b) particles.}
\label{3D-20}
\end{figure}

Stationary configurations of  small systems exhibit circular (2D) or spherical 
(3D) shell structure shown in \fref{2D-16} and \fref{3D-20}, respectively. 
Two-dimensional clusters are fully characterized by the occupation numbers of 
each shell. In 3D systems, however, there may be several states with slightly 
different energies having the same shell occupation numbers. To distinguish 
between such states, an intra-shell structure has to be specified, for example, 
by using the Voronoi symmetry parameter \cite{Ludwig}. For a larger number of 
particles, the nearly circular outer rings are gradually replaced by a distorted 
triangular lattice at the core of the cluster. As the screening increases 
and the inter-particle interaction range is thereby reduced, different shell 
configurations become stable and the size of clusters tends to shrink. For a
very strong screening, the interaction becomes extremely short-ranged, and 
particles cluster near the centre of the system where the confinement vanishes 
\cite{Candido}. As a result, an ordered triangular lattice is formed.

The number of stable configurations grows very rapidly with the number of
particles and reaches hundreds already for the moderate systems containing 
only several tens of particles. In order to describe this situation in a
quantitative manner it is often argued that \emph{the number of metastable
configurations grows exponentially with the system size}. However, as we
demonstrate in the present paper, the total number of stationary configurations
is not at all a useful concept -- especially for large systems -- and needs to
be refined. The reason is that different stable states are typically realized
with very different probabilities of occurrence, and consequently, many of
configurations, characterized by the smallest probabilities, remain
unnoticed within the finite duration of an experiment or a simulation.

Recently, an experimental study focusing on the occurrence probabilities of
the ground and metastable states in spherical clusters has been carried out
\cite{Kahlert}. The results indicate that sometimes metastable states are
formed with higher probabilities than the ground state. This happens, when
a metastable state controls a larger basin of attraction in the multidimensional
configuration space. The effect of the temperature, screening and the cooling 
rate \cite{Kahlert} on the probabilities of metastable states has been 
analyzed, and it was shown that stronger screening leads to higher probability 
of states with more particles on the inner shells. The authors also 
demonstrated, that the probabilities do not depend on the cooling rate when 
damping is strong.

The main goal of our work is to investigate the number and probabilities of
metastable states in systems of up to $100$ particles moving in two or three
dimensions and evaluate the configurational uncertainty. We introduce the 
concept of \emph{configurational entropy} as an objective measure of 
uncertainty, and investigate its dependence on the system size, dimensionality 
and screening. Thus cases of both Coulomb and Yukawa (screened Coulomb) 
inter-particle interactions are studied. The Coulomb forces are intended to 
represent the interaction between charged particles, e.~g., in arrays of 
electrons on the surface of liquid helium or semiconductor quantum dots, while 
the Yukawa potential models interactions in the complex (dusty) plasma 
environment \cite{Bonitz review}.

\section{Model system and computational approach}
\label{sec:model}

The model system consists of $N \leq 100$ identical particles, interacting 
through the Yukawa potential. Particles are allowed to move in two or three 
spatial dimensions and are confined by a symmetric parabolic potential with 
the characteristic confining frequency $\omega_0$:
\begin{equation}
 V_{\rm ext} (\mathbf{r})=\frac{1}{2} m\omega_0^2\mathbf{r}^2.
\end{equation}
The total potential energy of the system is given by
\begin{equation}
\label{eq:potential}
  U(\mathbf{r}_1, \ldots, \mathbf{r}_N)
  = \sum_{i=1}^N V_{\rm ext} (\mathbf{r}_i)
  + \sum_{i>j}^N \frac{Q^2}{r_{ij}}\rme^{-\kappa r_{ij}}.
\end{equation}
Here, $Q$ denotes the particle charge, 
$r_{ij}=\vert \mathbf{r}_i - \mathbf{r}_j\vert$ is the inter-particle distance,
and $\kappa$ stands for the screening parameter (the inverse screening length).
In the limit of $\kappa=0$, the interaction reduces to the pure Coulomb
potential. The two terms of \eref{eq:potential} are obviously in 
competition: The screened Coulomb repulsion tends to increase the distance 
between the particles, while the confinement tries to hold them together. By 
introducing the units of length $r_0=({Q^2}/{m\omega_0^2})^{1/3}$ and energy 
$E_0={Q^2}/{r_0}$, we can rewrite the potential energy \eref{eq:potential} in 
a simple dimensionless form
\begin{equation}
\label{eq:dimlesspotential}
  U(\mathbf{r}_1, \ldots, \mathbf{r}_N)
  = \sum_{i=1}^N \frac{1}{2}\mathbf{r}_i^2
  + \sum_{i>j}^N \frac{1}{r_{ij}} \rme^{-\kappa r_{ij}},
\end{equation}
where the constant $\kappa$ is now measured in the inverse $r_0$ units. The 
ground and metastable states of an interacting cluster correspond, respectively, 
to the global and local minima of \eref{eq:dimlesspotential}.

Our goal is to find the stationary states as well as the probabilities for the system to
settle in a given stationary state starting from a random
initial configuration. In order to give a precise and physically motivated
meaning to the above-mentioned randomness we choose to draw the starting
configurations from the canonical (Boltzmann) distribution. According to
this distribution, any initial configuration is in principle realizable,
however, its likelihood is proportional to $\exp({-E/kT})$. Here, $E$ stands for
the energy of the configuration and $kT$ is the simulation temperature, whose
role will be explained shortly.

Thus, the numerical procedure consists of two stages: First, a random
canonically distributed configuration of $N$ particles is created. Then, the
system is suddenly cooled and forced to roll down towards the closest energy 
minimum, or to be more precise, to the energy minimum whose basin of attraction
encloses the starting configuration.

The first stage is accomplished by employing the standard Metropolis
algorithm \cite{Metropolis} that generates a Markovian sequence of
configurations. Each new configuration is obtained from a previous one by a
random displacement of one of the particles. If the new configuration has a
lower energy it is accepted; otherwise the new state is accepted with
probability $\exp({-\Delta E / kT})$, where $\Delta E $ is the energy increment.
Higher values of temperature $kT$ allow the system to overcome potential
barriers between basins of attraction of different minima and thus efficiently
explore the configurational phase space.

Once the thermodynamic equilibrium is reached, the second stage of the algorithm
takes over. The system is suddenly cooled, that is, the closest local minimum
of the potential energy is located using downhill minimization techniques.
The phase space defined by all possible configurations of $N$ particles moving
in $D$ spatial dimensions is $(D\times N)$-dimensional, the distribution of its 
stationary points is quite complicated. Sometimes, the desired stationary point 
is located in a long and narrow curved valley. Therefore, it is not an easy 
task for any optimization algorithm to quickly locate the stationary states. In 
some cases, the steepest descent optimization technique has to be combined with
the Newton's method in order to increase the convergence rate and accuracy.

The whole heating and sudden cooling process is repeated a large number of
times (a typical number of simulation runs is $2\times 10^4$ -- 
$3\times 10^4$) to obtain accurate to one percent
probabilities of occurrence of different stable and metastable states. The
numerical experiment produces a list of found stable configurations and
their occurrence probabilities. This information --- while interesting on its
own --- may be further compressed into a single parameter that quantifies the
uncertainty due to the availability of multiple stable states. Following
the definitions accepted in information theory \cite{Jaynes} we define the
\emph{configurational entropy} as
\begin{equation}
  S = -\sum_{k=1}^M p_k \ln p_k.
\end{equation}
Here, $p_k$ is the probability of occurrence of $k$-th state, and $M$ is the
total number of found states. Performing a finite number of simulation runs
it is impossible to find all stable states for a system containing a large 
number of interacting particles. However, as we demonstrate in \sref{Results}, 
the contribution of such undiscovered configurations to the entropy is 
negligible, since their probabilities are very low. Hence the proposed 
configurational entropy is an objective measure of uncertainty of the 
particle configurations. It could also be interpreted as the logarithm of 
the \emph{effective} number of states. We note, that if there is only one 
available configuration, the entropy equals zero. This reflects the fact that 
we know the configuration that is to be found and thus there is no uncertainty.

\begin{figure}[ht]
\centering
\includegraphics[width=\figa]{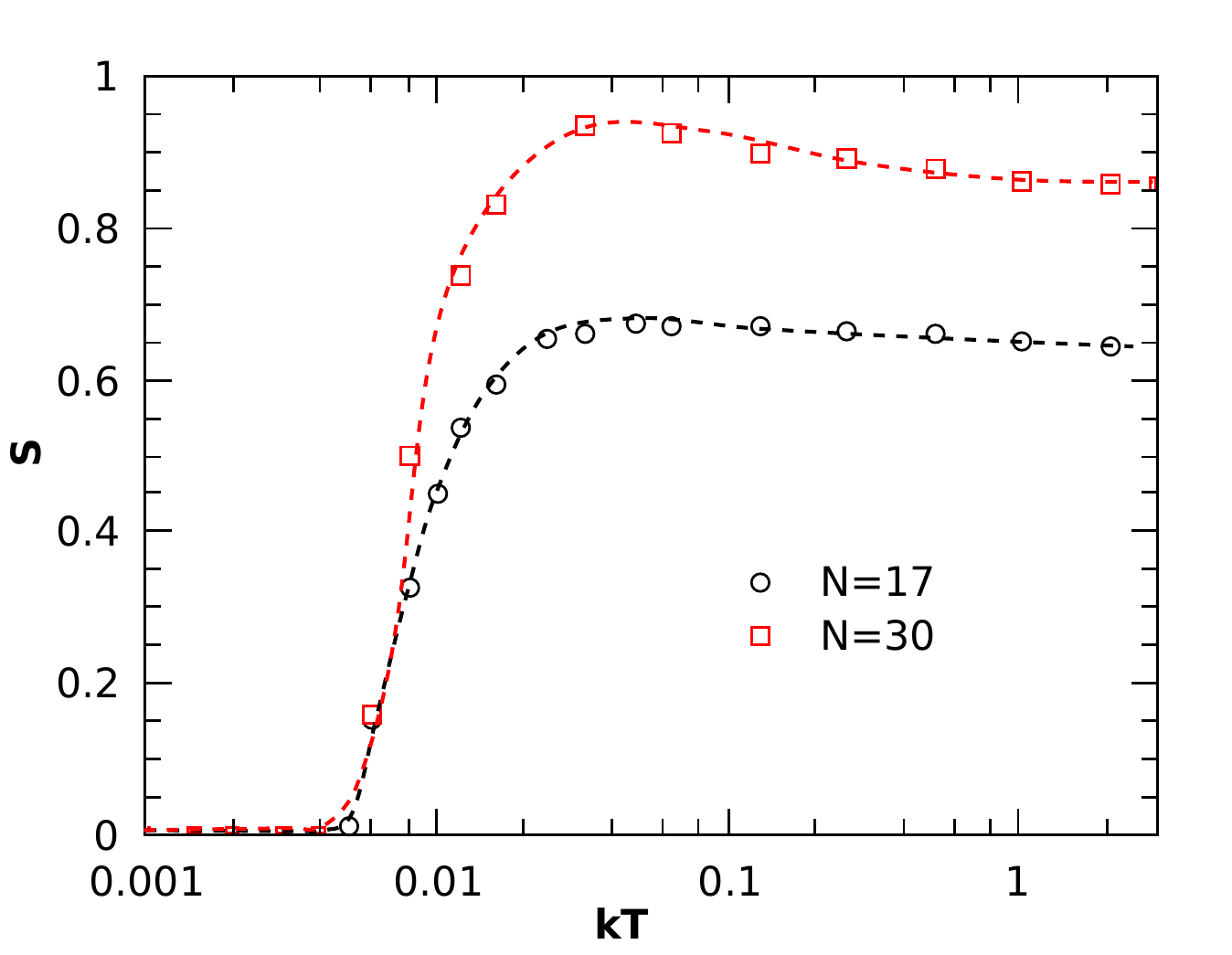}
\caption{Configurational entropy as a function of simulation temperature $kT$ 
for 2D systems with $17$ and $30$ particles and no screening ($\kappa=0$).}
\label{kT_entropy}
\end{figure}

With all that said, there remains an open question: The calculated probabilities
as well as the entropy will, in general, depend on the simulation temperature.
However, it is easy to see that these quantities change rapidly only at low
temperatures, that is, when the simulation temperature $kT$ is smaller than,
or comparable to, the heights of the potential barriers separating basins of
attraction of different minima. In the limit of high temperatures, the
obtained probabilities are essentially temperature independent. Therefore, 
in this limit the probabilities and the configurational entropy become
well-defined quantities.

These comments are illustrated in \fref{kT_entropy} which shows the dependence 
of the configurational entropy on the temperature of simulation. As long as 
the temperature $kT$ does not exceed a certain threshold value (around 
$kT \approx 0.005$ in this case), there is only one stationary configuration 
that the simulation would reveal -- the ground state. Naturally, in this limit 
there is no uncertainty and the configurational entropy remains close to zero. 
As the simulation temperature is increased, the potential barriers can be 
crossed and more states become available. Consequently, the calculated entropy 
rapidly increases. Eventually, in the high temperature limit, that sets in at 
$kT \approx 0.1$ in \fref{kT_entropy}, the system is able to reach the 
areas of attraction of any available energy minimum without limitations. Thus, 
the calculated configurational entropy levels off and becomes essentially 
temperature-independent. These observations confirm that the high-temperature 
limit of the entropy is indeed a well defined measure of uncertainty in the 
model system under investigation. The value of temperature used to determine
probabilities in our simulations is even higher ($kT = 2$).

\section{Results}
\label{Results}

We performed numerical simulations for 2D systems containing up to $N = 100$
particles at three values of the screening parameter $\kappa = 0$ (no 
screening), $\kappa = 1$ and $\kappa = 5$. In three dimensions we restricted
the numerically demanding calculations to two values of the screening strength
$\kappa = 0$ and $\kappa = 5$ and system sizes up to $N = 80$. Let us now turn 
to the obtained results. We start by discussing the number of stationary 
states, and then proceed to the behaviour of the configurational entropy.

\subsection{Number of metastable states}
\label{Number of metastable states}

\begin{figure}[ht]
\centering
\includegraphics[width=\figa]{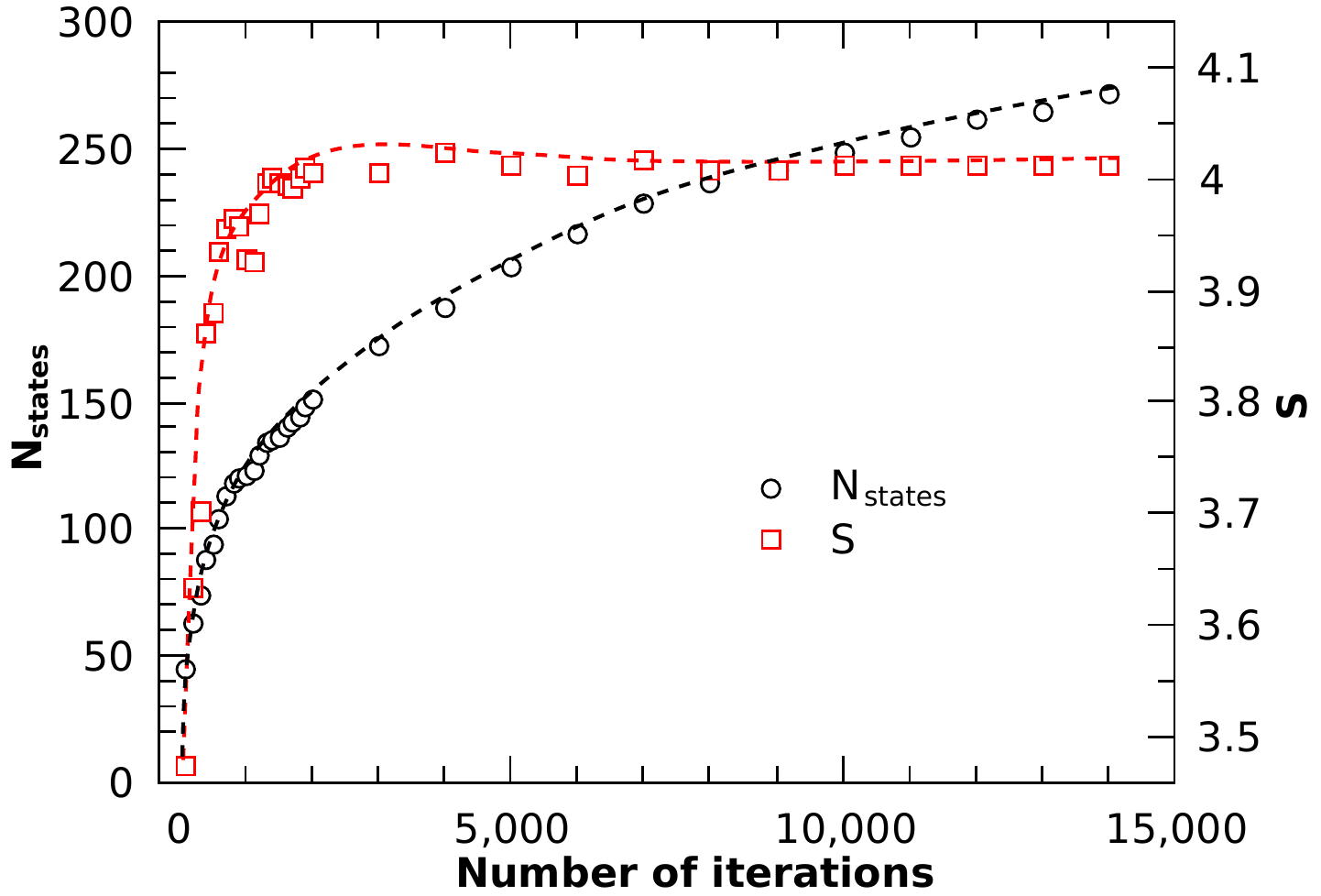}
\caption{Number of discovered stationary configurations $N_{\rm states}$ and
configurational entropy $S$, as a function of the number of performed
simulation runs for the unscreened 2D Coulomb cluster with $N=100$ particles.}
\label{iter_change}
\end{figure}

The computational procedure developed in \sref{sec:model} is capable of
determining all existing metastable configurations only for a small number
of particles $N$. However, when $N$ reaches a few dozen, even a relatively
large number of simulation runs does not guarantee the discovery of all 
stationary states. One just keeps discovering more and more stable 
configurations as the iterations are performed further.

\Fref{iter_change} shows the dependence of the number of discovered stable 
configurations $N_{\rm states}$ on the number of simulation runs for a 2D 
unscreened Coulomb cluster in a parabolic trap with $N=100$ particles. We see 
that the number of states grows steadily and fails to show any signs of 
convergence towards an asymptotic value as the number of simulation runs 
reaches $10\,000$. Evidently, a much larger number of iterations would be 
necessary to perform in order to achieve convergence. These observations 
confirm that, in this context, the notion of the total number of stable 
configurations is ill-defined, and consequently, is not that useful.

In contrast, the configurational entropy grows rapidly from the very beginning
and reaches its asymptotic value already after a few thousand of simulation
runs (see \fref{iter_change}). There obviously exist just a few most
important configurations with relatively high probabilities of occurrence and
a large number of states with insignificant probabilities. The latter make
only a minor contribution to the entropy, and its value does not change
appreciably as the number of discovered states increases continuously.
Therefore we may conclude, that the configurational entropy is, indeed, an 
objective parameter of uncertainty even when number of simulation runs is 
limited and only a fraction of all stable configurations is discovered.

\begin{figure}[ht]
\centering
\includegraphics[width=\figa]{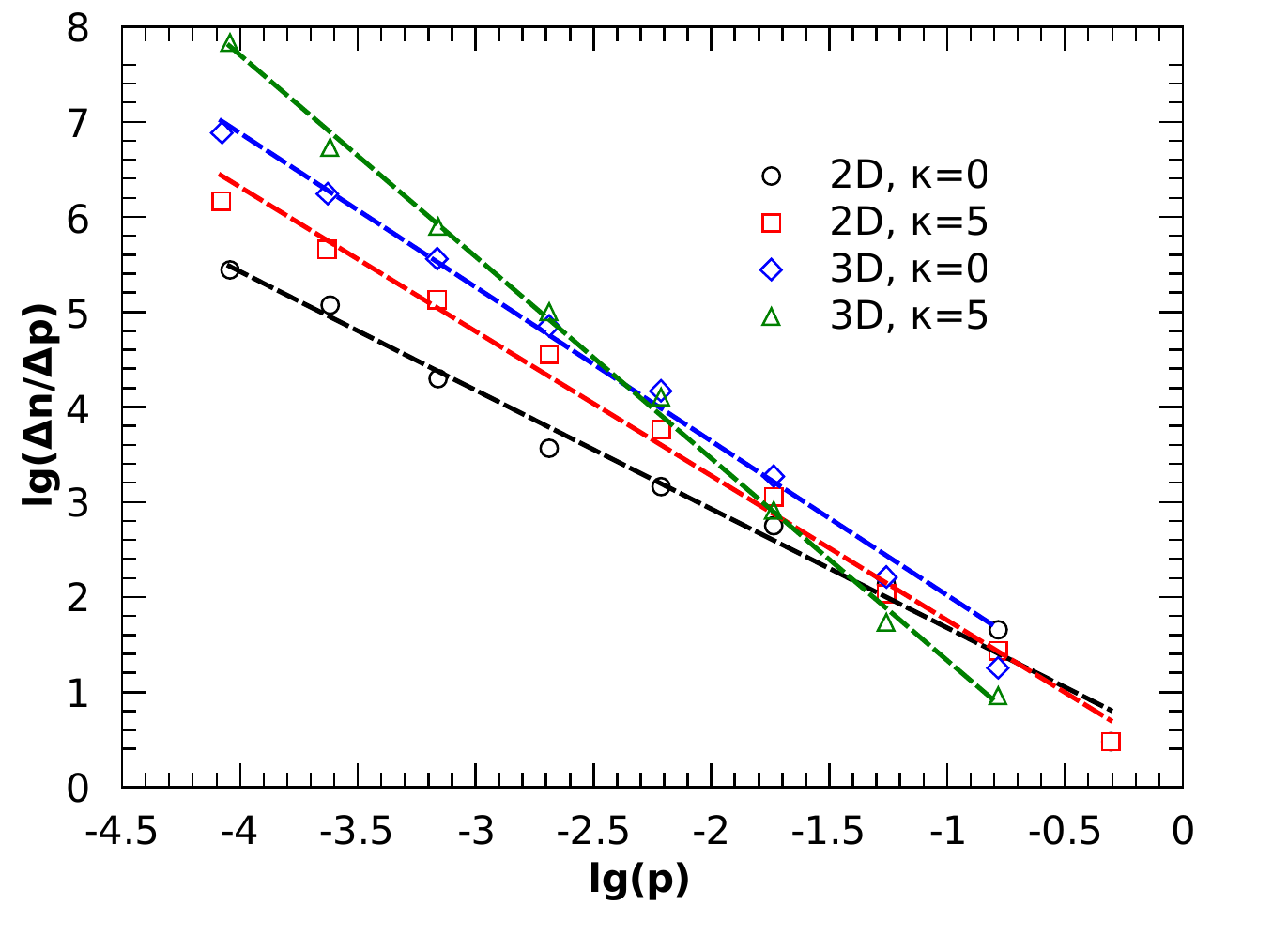}
\caption{Power-law dependence of the normalized number of states
$\Delta n/ \Delta p$ whose probabilities lie in the interval
$(p-\Delta p, \; p )$ on the value of $\lg (p)$ for a system with
$N=80$ particles.}
\label{PowLaw}
\end{figure}

As a matter of fact, the seemingly unrelenting growth of the number of
discovered states is due to a heavy-tailed power-law distribution of the
state probabilities. Let us investigate more closely the probabilities of the
discovered stationary states for the $80$-particle system. The range of
probabilities is divided into the intervals of width $\Delta p$ and the number of
states $\Delta n$ with probabilities of occurrence within the interval
$(p-\Delta p,\; p)$ is calculated. \Fref{PowLaw} shows that the
dependence of $\Delta n/\Delta p$ on the probability $p$ is linear on a 
log-log scale. That is,
\[
  \lg (\Delta n/ \Delta p)\approx - \alpha \lg (p) + \rm{const},
\]
which is equivalent to
\begin{equation}
  \Delta n/ \Delta p \sim p^{-\alpha}.
  \label{PowLaweq}
\end{equation}
Hence, there is only a tiny number of dominant configurations with
significantly high chance of occurrence. The number of discovered states in 
a given probability interval increases rapidly as the probabilities become 
lower. However, the areas of attraction of those numerous stationary states 
account only for a very small fraction of the total phase-space volume.

The distribution \eref{PowLaweq} follows a power law which is in various 
contexts known as the Zipf's law or Pareto distribution. It is worth noting, 
that power laws appear widely in physics, economics, computer science, biology, 
geography, demography and social sciences \cite{Newman}. Zipf's law is observed 
in various phenomena that have no definite characteristic scale and distribution 
of the relevant quantities spans many orders of magnitude. A few notable 
examples of Pareto-distributed items are the sizes of cities, magnitudes of 
earthquakes, frequencies of words in natural languages and so on.

We observe that the four lines shown in \fref{PowLaweq} have different
slopes; this corresponds to different values of the power-law exponents
$\alpha$. The lowest value of the exponent $\alpha$ seen in \fref{PowLaweq}
is found to be $1.25$ and characterizes the two-dimensional cluster with
unscreened Coulomb interaction. In general, one notes that for both 2D and 3D 
systems, the exponent $\alpha$ increases as the range of inter-particle 
interaction becomes shorter. Thus, we may conclude that \emph{screening leads
to the proliferation of low-probability states}.

The results obtained for systems with a different number of particles are
similar to those pertaining to $N = 80$ shown in \fref{PowLaw}. Of course,
we have to restrict the consideration to large systems, so that a sufficient
amount of data for statistical analysis is generated. The value of the exponent 
$\alpha$ is not universal and slightly grows with the system size. Its value 
for a 2D system with $\kappa=0$ reaches 1.4 for $N=100$.

\begin{figure}[ht]
\centering
\includegraphics[width=\figa]{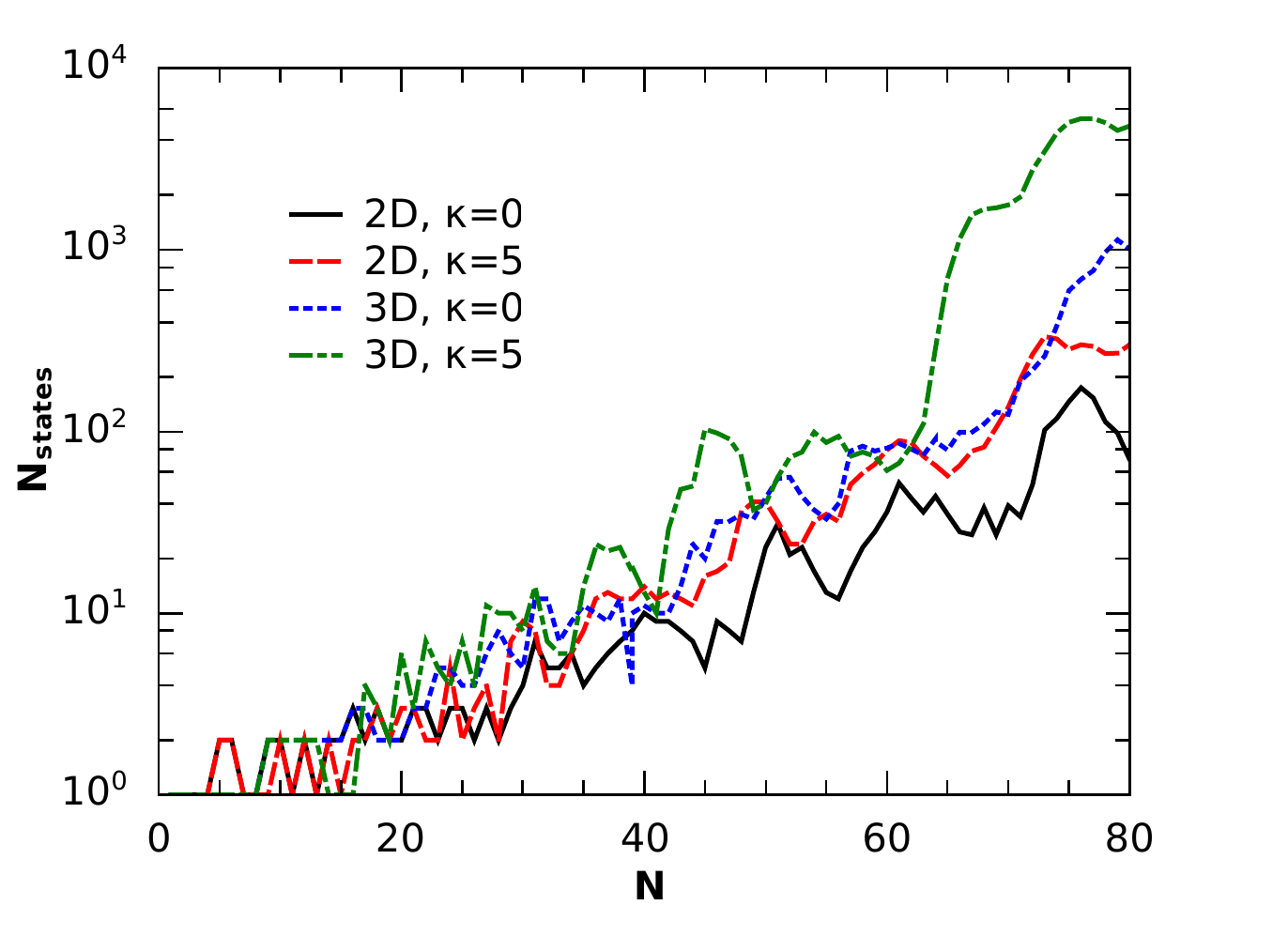}
\caption{The number of stationary states discovered after $3\times 10^4$
simulation runs in the two- and three-dimensional systems of $N$ particles 
with screening $\kappa$.}
\label{found-states}
\end{figure}

The number of discovered stationary states of the model system grows very 
rapidly with the number of particles. \Fref{found-states} shows the dependence
of the number of configurations, found after $30\,000$ simulation runs, 
on the number of particles $N$ for 2D and
3D systems without screening and with screening $\kappa=5$. On the average, the 
growth is nearly exponential (note the logarithmic scale on the vertical axis) 
and is faster for the systems with shorter range of inter-particle interaction. 
Just to give an example, we observe that in a 2D system with $\kappa=0$ the 
number of states does not exceed three as long as $N<16$. In contrast, in a 
system with the screening parameter $\kappa=20$, three stable configurations 
exist already when $N=10$.

Note, however, that here we are discussing the number of \emph{discovered}
states and not the true number of states, which is unknown. With this
reservation, we may conclude that our simulations confirm (or at least, do not 
reject) the oft-cited conjecture of an exponential growth of the number of 
states.

\subsection{Entropy in 2D and 3D systems}
\label{Entropy}

The configurational entropy, as expected, grows rapidly with the number of
particles. The growth is quite erratic, with conspicuous maxima and minima.

\begin{figure}[ht]
\centering
\includegraphics[width=\figa]{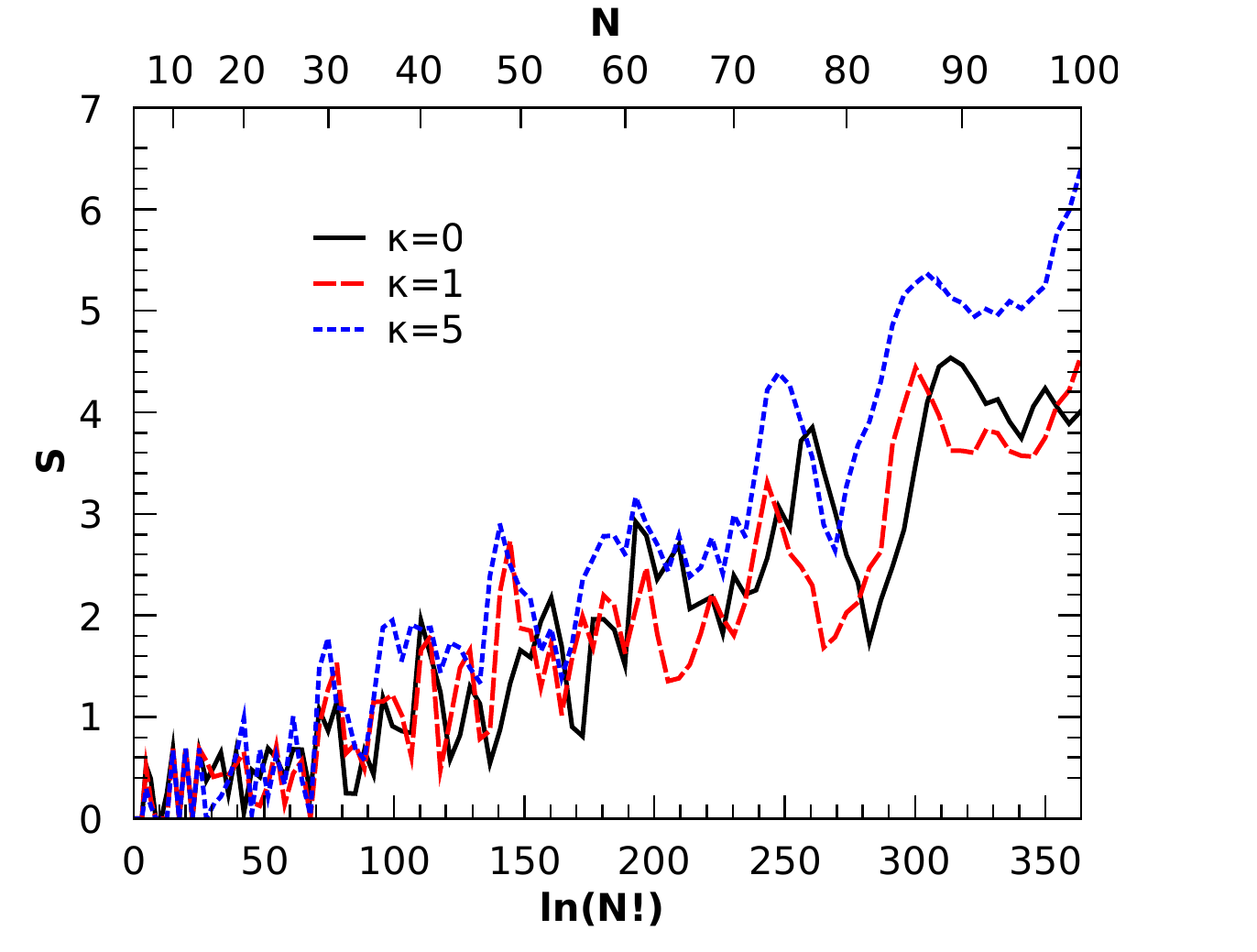}
\caption{Configurational entropy of a 2D system with screening parameter
$\kappa$ as a function of parameter $\ln (N!)$.}
\label{entropy}
\end{figure}

\begin{figure}[ht]
\centering
\includegraphics[width=\figa]{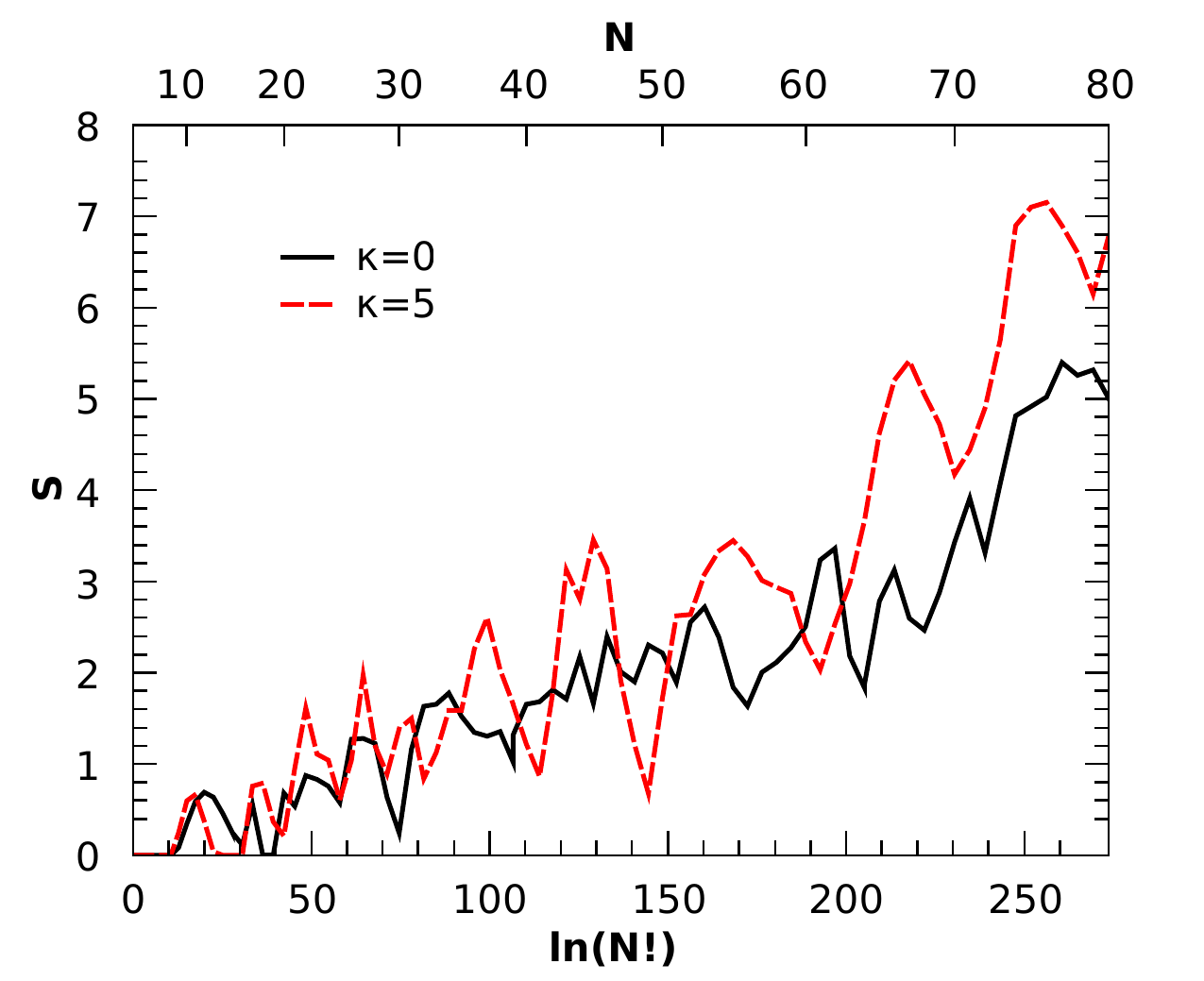}
\caption{Configurational entropy of a 3D system with screening parameter
$\kappa$ as a function of parameter $\ln (N!)$.}
\label{entropy3D}
\end{figure}

On the average, the dependence of the entropy on the number of particles in 
both 2D and 3D systems with different values of screening parameter $\kappa$ 
is clearly faster than linear. That is, the effective number of states
$\Omega (N)=\exp[{S(N)}]$ grows faster than exponentially. In view of the
combinatorial nature of the problem, it is tempting to hypothesize that the 
effective number of states could be proportional to the factorial of the 
number of particles $N!$, and the entropy behaves as 
\begin{equation}
\label{eq:hypothesis}
  S = \gamma \ln(N!).
\end{equation}
Figures \ref{entropy} and \ref{entropy3D} show the calculated dependence of 
the entropy on the argument $\ln{(N!)}$ for 2D and 3D clusters with different 
screening lengths. For the sake of convenience, these figures are also supplied
with a nonlinear top axis that gives the corresponding values of $N$. We see,
that the averaged trends indeed exhibit a linear growth. While it is impossible 
to rigorously ``prove'' conjectures of this sort, the results of the numerical 
experiment clearly do not contradict \eref{eq:hypothesis}. Thus, taking
advantage of this conjecture we may use the values of the slope $\gamma$ as 
a convenient measure of the entropy growth. The values of $\gamma$ for the 2D 
and 3D crystals with different values of $\kappa$ are collected in 
\tref{Table:Gammas}.

\Table{\label{Table:Gammas}Values of the slope $\gamma$ for 2D and 3D systems 
with different values of screening constant $\kappa$.}
\begin{tabular}{@{}llll}
\br
	 Dimension& $\kappa=0$ & $\kappa=1$ & $\kappa=5$\\
\mr
	2D      & 0.011 & 0.011 & 0.015\\
	3D      & 0.015 & --- & 0.021\\
\br
\end{tabular}
\endTable

For the systems with small screening, the peak value of the entropy is close to 
$S = \ln 2$ as long as the number of particles is lower than $N=20$. There are
only up to $3$ stable configurations discovered, and the entropy is close to
$\ln 2$ when mostly two of them have comparable probabilities of occurrence. 
A typical example is a 2D cluster with $N=14$, entropy $S=0.69$ and two
configurations with probabilities $0.44$ and $0.56$.

As the number of particles increases, the areas of attraction of many 
configurations become comparable in size and, as a consequence, the entropy 
grows. As an example, one of the peaks of the configurational entropy in a 
system with $\kappa=0$ is located at $N=40$. There are ten stable states, only 
three of them with high, and comparable, probabilities of occurrence 
($0.23,\; 0.23,\; 0.22$), the remaining ones being lower by an order of 
magnitude or more.

On the other hand, the uncertainty is low when there is only one (or very few, 
in comparison to the total number of states) dominant configurations whose
probabilities are much higher than the remaining ones. The entropy of the 3D 
system with $\kappa=5$ has a sharp minimum at $N=49$. There are $37$ discovered
stationary states, however, only one of them has a high chance of occurrence
($0.85$).

Although the number of discovered states depends on interaction range
(see \fref{found-states}), screening only weakly affects the configurational
entropy of small 2D systems.
Even more, the positions of  major falls and growths of the entropy for
different values of screening parameter $\kappa$ are close to each other.
The most noticeable changes in uncertainty of the system occur near
$N\approx 30, \; 40, \;48, \; 74, \; 87$, where the addition of a few
particles firstly almost doubles the value of entropy, and then suddenly
decreases it to the much lower value.

Slightly different situation is observed in three dimensions 
(\fref{entropy3D}). Configurational entropy on the average grows
faster ($\gamma=0.015$ for $\kappa=0$). Even for a small number of particles,
increased screening leads to the sharper and higher peaks of the
uncertainty. The most noticeable maxima of the entropy in a 3D Yukawa
crystal with $\kappa=5$ occur at $N=27,\; 37,\; 45,\; 67,\; 76$. 
Growth of the entropy in a Coulomb system ($\kappa=0$) is much smoother, the 
number of peaks is higher. Also, as opposed to the 2D system, no
relations can be found between the positions of peaks for the different
values of screening constant $\kappa$.

\begin{figure}[ht]
\centering
\includegraphics[width=\figa]{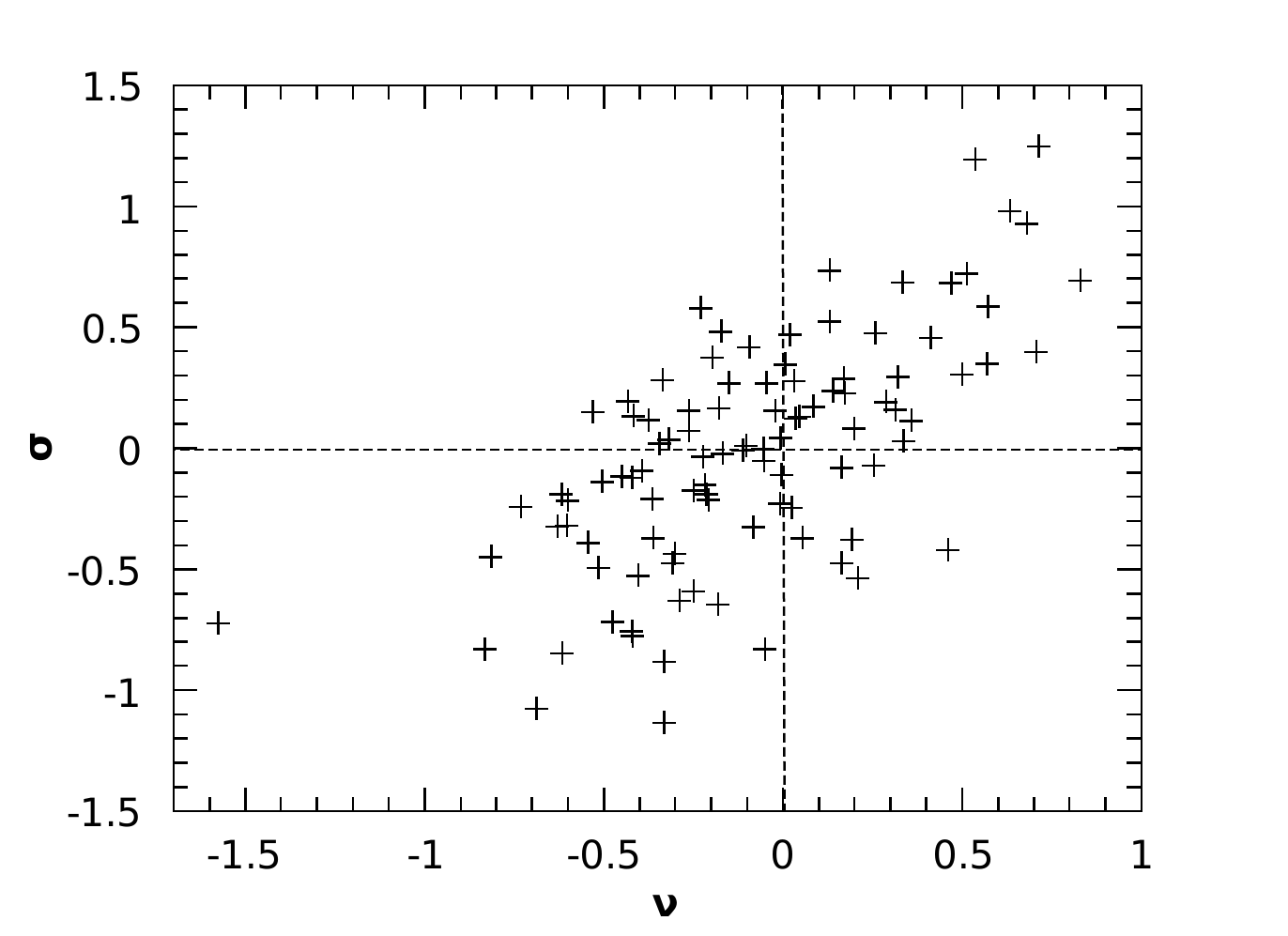}
\caption{Correlations between fluctuations of entropy $\sigma$ and the 
logarithm of the number of discovered stable configurations $\nu$ in a 
2D system with screening constant $\kappa=1$.}
\label{correlations}
\end{figure}

It is worth mentioning that although only a fraction of all possible stable 
configurations is found during our simulations, there exists a correlation 
between the number of discovered states and the effective number of 
configurations $\Omega=\exp(S)$. As the number of particles grows, the 
increasing number of discovered states parallels the growth of the entropy 
and vice versa. To illustrate this point, we analyze the behaviour of the 
logarithm of the number of states and entropy in a two-dimensional system 
with $\kappa=1$. The entropy $S$ and the logarithm of $N_{\rm states}$ are 
separated into the sums of the main linear part and the fluctuating term:
\begin{eqnarray}
  S = \gamma \ln(N!) + \sigma (N),\\
  \ln(N_{\rm states}) = \beta N + \nu (N).
\end{eqnarray}
The correlation plot between the fluctuating terms $\sigma$ and $\nu$ for 
all values of $N \leq 100$ is shown in \fref{correlations}. We see, that 
nearly all points (with the exception of a single outlier) fall inside an 
elongated elliptic area, thus confirming the presence of correlations. 
Similar results are obtained for the systems with $\kappa=0$ and $\kappa=5$. 
In three dimensions, however, the correlation between the number of discovered 
states and the configurational entropy is weaker. 

\section{Conclusions}
In conclusion, using the Monte Carlo and downhill minimization algorithms 
we have investigated the probabilities of metastable states of two- and 
three-dimensional classical particle clusters confined by a parabolic 
potential trap. It is demonstrated, that the total number of states is not a 
useful parameter to describe large systems due to its sensitivity to the low 
probabilities of the majority of stable configurations. The number of states 
with low chance of occurrence is further increased when the inter-particle 
potential is short-ranged. In order to quantitatively evaluate the uncertainty 
due to the presence multiple stable states, we propose to rely on the concept 
of the configurational entropy -- a  quantity that accurately describes the 
randomness of the final state even when the number of simulation runs is 
limited. It turns out that the effective number of states $\exp(S)$ grows 
faster than exponentially, that is, on the average $S = \gamma \ln(N!)$ for 
$N<100$. Entropy grows faster in the systems with short inter-particle 
interaction range.

\ack
Ar\={u}nas Radzvilavi\v{c}ius acknowledges Student Research Fellowship Award 
from the Lithuanian Science Council.

\end{document}